\documentclass[letterpaper,twoside]{article}
\usepackage{amssymb}
\usepackage{amsmath}
\usepackage{latexsym}
\usepackage{verbatim}
\usepackage{epsfig}
\usepackage{fancyhdr}
\usepackage[english]{babel}
\newcommand{\papertitle}{%
Unstable particles in non-relativistic quantum\\ mechanics?}
\newcommand{\runningtitle}{%
Unstable particles in non-relativistic quantum mechanics?}
\newcommand{\pauthor}{%
H.{} Hernandez-Coronado
}
\newcommand{\paperauthor}{%
H.{} Hernandez-Coronado%
}
\begin{document}
%\initfloatingfigs
%%%%%%%%%%%%%%%%%%%%%%%%%%%%%%%%%%%%%%%%%%%%%%%%%%%%%%%%%%%%%%
%%%%%%%%%%%%%%%%%%%%%%%%%%%%%%%%%%%%%%%%%%%%%%%%%%%%%%%%%%%%%%
%%%%%%%%%%%%%%%%%%%%%%%%%%%%%%%%%%%%%%%%%%%%%%%%%%%%%%%%%%%%%%
% Titlepage
%%%%%%%%%%%%%%%%%%%%%%%%%%%%%%%%%%%%%%%%%%%%%%%%%%%%%%%%%%%%%%
\begin{titlepage}
\vspace*{-1cm}
\begin{flushright}
\textsf{}
%\\
%\textsf{ICN-UNAM-yy/pp}
\\
\mbox{}
\\

\end{flushright}
%%%%%%%%%%%%%%%%%%%%%%%%%%%%%%%%%%%%%%%%%%%%%%%%%%%%%%%%%%%%%%
%%%%%%%%%%%%%%%%%%%%%%%%%%%%%%%%%%%%%%%%%%%%%%%%%%%%%%%%%%%%%%
%%% TITLE, AUTHORS
%%%%%%%%%%%%%%%%%%%%%%%%%%%%%%%%%%%%%%%%%%%%%%%%%%%%%%%%%%%%%%
%%%%%%%%%%%%%%%%%%%%%%%%%%%%%%%%%%%%%%%%%%%%%%%%%%%%%%%%%%%%%%
% \begin{center}
%%%%%%%%%%%%%%%%%%%%%%%%%%%%%%%%%%%%%%%%%%%%%%%%%%%%%%%%%%%%%%
\renewcommand{\thefootnote}{\fnsymbol{footnote}}

\begin{flushright}
\textsf{\today}
\end{flushright}
\vspace{3cm}

\begin{LARGE}
\noindent\bfseries{\sffamily \papertitle}
\end{LARGE}

\noindent \rule{\textwidth}{.5mm}

\vspace*{1.6cm}

\noindent 
\begin{large}%
\textsf{\bfseries\pauthor}
\end{large}

\vspace*{.1cm}

\phantom{XX}
\begin{minipage}{1\textwidth}
\begin{it}
\noindent 
Instituto Mexicano del Petr\'oleo\\
Eje central L\'azaro C\'ardenas 152, 07730, M\'exico D.F., M\'exico.\\
\end{it}
\texttt{hhernand@imp.mx
\phantom{X}}
\end{minipage}

\vspace*{3cm}
%%%%%%%%%%%%%%%%%%%%%%%%%%%%%%%%%%%%%%%%%%%%%%%%%%%%%%%%%%%%%%
%%% ABSTRACT
%%%%%%%%%%%%%%%%%%%%%%%%%%%%%%%%%%%%%%%%%%%%%%%%%%%%%%%%%%%%%%
\noindent
\textsc{\large Abstract:}
The Schr\"odinger equation is up-to-a-phase invariant under the Galilei group. This phase leads to the Bargmann's superselection rule, which forbids the existence of the superposition of states with different masses and implies that unstable particles cannot be described consistently in non-relativistic quantum mechanics. In this paper we claim that Bargmann's rule neglects physical effects and that a proper description of non-relativistic quantum mechanics requires to take into account this phase through the Extended Galilei group and the definition of its action on spacetime coordinates.

\end{titlepage}
\setcounter{footnote}{0}
\renewcommand{\thefootnote}{\arabic{footnote}}
\setcounter{page}{2}
% %%%%%%%%%%%%%%%%%%%%%%%%%%%%%%%%%%%%%%%%%%%%%%%%%%%%%%%%%%%%%%
% %%%%%%%%%%%%%%%%%%%%%%%%%%%%%%%%%%%%%%%%%%%%%%%%%%%%%%%%%%%%%%
% %%%%%%%%%%%%%%%%%%%%%%%%%%%%%%%%%%%%%%%%%%%%%%%%%%%%%%%%%%%%%%
%\noindent \rule{\textwidth}{.5mm}
%
%%\tableofcontents
%
%\noindent \rule{\textwidth}{.5mm}\\
\section{Introduction}

The Schr\"odinger equation is not invariant under pure Galilean boosts, but up-to-a-phase invariant. This implies that there is not a unitary representation of the Galilei group in the Hilbert space of a non-relativistic particle, but a projective representation\footnote[1]{In general, for a projective representation, two elements $g,g'$ of the Galilei group are represented in the Hilbert space by operators $U_g$ and $U_{g'}$ such that $U_g U_{g'}=\omega(g,g')U_{g\cdot g'}$, with $|\omega|=1$.} \cite{Az,LL}. The Galilei projective representation can be understood as a unitary representation of the Extended Galilei group, a central extension of the Galilei group \cite{Az,LL} and it yields to the Bargmann's superselection rule, {\it i.e.}, the impossibility of consistently describing superposition of states with different mass in non-relativistic quantum mechanics \cite{Bar}. 

However, the phase $\omega$ (see footnote) associated with the mentioned projective representation describes physical effects, {\it i.e.}, it is related to relativistic time differences between the corresponding inertial observers to order $1/c^2$. Moreover, as pointed out by Greenberger \cite{Gren}, there are physical situations that the superselection rule neglects, {\it e.g.}, binding energy effects which ``show up as inertial mass when one transforms to another system [of reference]''. Also, Giulini has remarked that for the mass to define a superselection rule, it has to have spectrum and then, it has to be dynamical and the conjugate momentum of a canonical coordinate \cite{Giu95}. 

The situation is, as Anandan put it, that the symmetry group of non-relativistic quantum mechanics (Extended Galilei group) is different to the spacetime's one (Galilei group) \cite{Ana}. And the Bargmann's rule arises from assuming that the symmetry group describing the world in the low velocity limit is the Galilei group and the phase associated with the Extended group is unphysical.

In this paper we claim, echoing previous works \cite{Gren, Mash, HB, Gren83}, that the phase $\omega$ is physically meaningful, we provide as an example of which the Sagnac effect and contrast it with the argument behind the Bargmann's rule. Thus, we propose that a consistent description of non-relativistic quantum mechanics requires to realize that the symmetry of the physical world in the non-relativistic limit is given by the extended Galilei group, and then that we should define an action of it on the coordinates of spacetime consistent with its representation in the corresponding Hilbert space.

\section{Bargmann's superselection rule}

It is well known that the Schr\"odinger equation is invariant under Galilean transformations up-to-a-phase. To see this, let us assume that the Schr\"odinger equation in the presence of a scalar potential $V$ is valid in an inertial reference frame $S$
with coordinates ($\mathbf{x},t$), {\it i.e.},
\begin{equation}
 i\hbar\partial_t\psi(\mathbf{x},t)=\left(mc^2-\frac{\hbar^2}{2
m}\nabla^2+V(\mathbf{x},t)\right)\psi(\mathbf{x},t),\label{SchEq}
\end{equation}
and then, let us define another inertial reference frame $S'$ with coordinates $(\mathbf{x}',t')$ related to the unprimed ones by the following Galilean transformation, $\mathbf{x'}=\mathbf{x}-\mathbf{v} t$ and $t'=t$, which imply that $\partial_{t}'=\partial_{t}+\mathbf{v}\cdot\nabla$ and $\nabla'=\nabla$, where $\mathbf{v}$ is a constant three dimensional real vector. In the frame $S'$, Eq. (\ref{SchEq}) can be written as:
\begin{equation}
 i\hbar\partial_{t}'\psi'(\mathbf{x'},t')=\left(mc^2-\frac{\hbar^2}{2
m}\nabla'^2+V'(\mathbf{x'},t')\right)\psi'(\mathbf{x'},t'),\label{SchEqp}
\end{equation}
with
\begin{equation}
\psi'(\mathbf{x'},t')=e^{im(\mathbf{v}^2t/2-\mathbf{v}\cdot\mathbf{x})/\hbar}
\psi(\mathbf{x},t),\label{psip}
\end{equation}
and $V'(\mathbf{x'},t')=V(\mathbf{x},t)$. The relation (\ref{psip}) is the standard transformation expression for the wavefunction under a Galilean boost and it implies that the probability density is a Galilean scalar when $m\in\mathbb{R}$ although the wavefunction is not. As it is also well known, under a space translation by $\mathbf{a}\in\mathbb{R}^3$ the wavefunction transforms as $\psi_{\mathbf{a}}(\mathbf{x}+\mathbf{a},t)=\psi(\mathbf{x},t)$, where $\psi_{\mathbf{a}}=e^{-i\mathbf{P}\cdot\mathbf{a}/\hbar}\psi$ with $P_i$ being the spacial translation generator along the $i$-th coordinate\footnote{Latin indexes label spacial coordinates, {\it i.e.}, $i,j=1,2,3$.}.

The relation (\ref{psip}) traduces in that the representation of the Galilei group in the Hilbert space of a non-relativistic particle is not a true unitary representation but a projective one. Projective representations are physically relevant because pure states are represented by rays in the Hilbert space.

As a consequence of the above mentioned projective representation arises the Bargmann's superselection rule, {\it i.e.}, the impossibility of describing superposition of states with different mass in a consistent way. In order to see this, we will reproduce here the original Bargmann's argument \cite{Bar}. Suppose $\psi(\mathbf{x},t)=\psi_{m_1}(\mathbf{x},t)+\psi_{m_2}(\mathbf{x},t)$ is a superposition of the states with two masses in an inertial reference frame $S$. Now let us perform the following composed transformation: a translation (by $\mathbf{a}$) and then a boost (by $\mathbf{v}$) followed by a translation (by $-\mathbf{a}$) and finally a boost (by $-\mathbf{v}$) to return to the original inertial system $S$ with coordinates $(\mathbf{x},t)$. In the light of the transformation laws for the wavefunction under boosts (\ref{psip}) and spacial translations, it is not difficult to see that under the previous composed transformation the wavefunction $\psi(\mathbf{x},t)$ yields:
\begin{equation}
\tilde{\psi}(\mathbf{x},t)=e^{im_1 \mathbf{v}\cdot\mathbf{a}/\hbar}\psi_{m_1}(\mathbf{x},t)+e^{im_2 \mathbf{v}\cdot\mathbf{a}/\hbar}\psi_{m_2}(\mathbf{x},t),\label{psipp}
\end{equation}
which is equal to $\psi(\mathbf{x},t)$ (modulo a phase) only if the Bargmann's superselection rule is imposed, {\it i.e.}, $m_2=m_1$. 

\subsection{Relativistic remnants}

The phase appearing in expression (\ref{psip}) and giving rise to the Bargmann's rule, however, has a physical origin. To understand where it comes from it is required to look into the relativistic case \cite{Mash, HB, Gren83}. The Klein-Gordon equation, 
\begin{equation}
\left(\frac{1}{c^2}\frac{\partial^2}{\partial t^2}-\nabla^2+\frac{m^2 c^2}{\hbar^2}\right)\phi(x)=0,\label{KG}
\end{equation}
with the ansatz:
\begin{equation}
\phi(x)=e^{-imc^2 t/\hbar}\psi(\mathbf{x},t),\label{Rwf}
\end{equation}
reduces to:
\begin{equation}
-\frac{\hbar^2}{2m}\nabla^2\psi(\mathbf{x},t)=i\hbar\partial_t \psi(\mathbf{x},t)-\frac{\hbar^2}{2mc^2}\partial_t^2\psi(\mathbf{x},t),\label{KGlimEq}
\end{equation}
and then by neglecting the last term in the r.h.s. of the above equation in the limit $c\rightarrow\infty$, the Schr\"odinger equation is obtained. Now, since $\phi$ is a scalar, it follows from (\ref{Rwf}) that the wavefunctions for two Lorentz inertial observers are related by:
\[
\psi'(\mathbf{x}',t')=e^{imc^2(t'-t)/\hbar}\psi(\mathbf{x},t)=e^{im(\mathbf{v}^2t/2-\mathbf{v}\cdot\mathbf{x}) /\hbar}\psi(\mathbf{x},t)+\mathcal{O}(1/c^2),
\]
such that in the non-relativistic limit, the previous expression reduces precisely to relation (\ref{psip}). Thus, the phase in relation (\ref{psip}) can be identified as some relativistic remnant: it is the time difference of order $1/c^2$ between the different inertial observers.

Consequently, by imposing the Bargmann's superselection rule we may be neglecting physical effects. We claim that this is indeed the case. In fact, it has been suggested that the phase shift observed in the COW experiment \cite{Gren83} as well as the Sagnac effect \cite{Ana, Die90} may lead to such a situation (for the experimental measurement of such effects see \cite{cow, csw}, respectively). We will focus in the latter effect in the following subsection.

\subsection{Sagnac effect}

Consider a device (source/detector) mounted on the extreme of a disk of radius $R$. Suppose that the device can emit two equal signals restricted to move along circular paths whose center is the disk's, with the same velocity and in opposite directions. Now, the disk is set into rotation around its center with angular velocity $\Omega$. We would like to know if there is a detectable physical effect due to the rotation when they arrived back to the emission point. We can describe the previous system from both, the non-relativistic and relativistic frameworks and they produce different results \cite{Ana, Die89}. 

Non-relativistically, both signals arrive at the same time at the detector and if the signals are Galilean invariant waves, their phases are equal such that there is not interference. This result can be obtained by an observer in a laboratory frame and also by a comoving observer (which can be thought of as an instantaneous inertial Galilean observer). 

Relativistically, on the other hand, the two signals do not arrive at the same time, but with a difference $\Delta t=4\pi R^2\Omega/\sqrt{c^2-\Omega^2R^2}$ (as measured by a comoving observer) \cite{Die89}. This time difference traduces in the phase difference $\Delta \phi=4\pi R^2\Omega\omega/\sqrt{c^2-\Omega^2R^2}$, where $\omega$ is the signal frequency. Again, this result can be obtained by an observer in a laboratory frame and also by a comoving observer (which can be thought of as an instantaneous inertial Lorentzian observer). By using the Einstein-Planck relation $\hbar\omega=mc^2/\sqrt{1-v^2/c^2}$ and taking the non-relativistic limit, the above phase difference reduces to $\Delta\phi_N=4\pi R^2 m\Omega/\hbar$.

Now, quantum mechanically, the signals' phases transform according to the projective representation (\ref{psip}) such that its difference corresponds to
\[
\Delta \phi_{NQM}=m(v+\Omega R)^2 t/2\hbar-m(v-\Omega R)^2 t/2\hbar=\Delta\phi_N,
\]
with $t=\pi/\Omega$. 

Consequently, if a Sagnac-like experiment is performed with a superposition of states with different masses $\psi=\psi_{m_1}+\psi_{m_2}$, we would expect to obtain an interference term similar to the one obtained from the Bargmann's superselection rule (\ref{psipp}) and we know that it makes sense physically, so then we do not have to impose any superselection rule.

\section{Extended Galilei group}

The projective representation of the Galilei group we found is equivalent to a true unitary representation of the extended Galilei group \cite{Az, LL, Ana}, which consists of introducing another generator $M$ that commutes with all other Galilei group generators and such that the corresponding algebra remains the same except for the relation $[C_i,P_j]= M\delta_{ij}$, where $C_i$ is the Galilean boost generator along the $i$-th coordinate. In terms of this group, the Bargmann composed transformation is represented by 
\begin{equation}
e^{-i\mathbf{v}\cdot\mathbf{C}/\hbar} e^{-i\mathbf{P}\cdot\mathbf{a}/\hbar} e^{i\mathbf{v}\cdot\mathbf{C}/\hbar} e^{i\mathbf{P}\cdot\mathbf{a}/\hbar}=e^{iM\mathbf{a}\cdot\mathbf{v}/\hbar},\label{EGTBT}
\end{equation}
which is not the group identity (as was for the Galilei group). 

In order to see that the previous result is physically meaningful, let us compare with the relativistic case. In the Poincar\'e algebra the commutator between the boost generators $\mathbf{K}$ and the spacial translations $\mathbf{P}$ is given by $[K_i,P_j]=H\delta_{ij}/c^2$, which reduces precisely to the commutator between $C_i$ and $P_j$ to leading order in the non-relativistic limit. Correspondingly, the relativistic version of the transformation considered by Bargmann can be written, up to $\mathcal{O}(1/c^2)$, as:
\begin{equation}
 e^{-i \mathbf{v}\cdot\mathbf{K}/\hbar}e^{-i\mathbf{a}\cdot\mathbf{P}/\hbar}e^{i \mathbf{v}\cdot\mathbf{K}/\hbar}e^{i\mathbf{a}\cdot\mathbf{P}/\hbar}=e^{iH \mathbf{a}\cdot\mathbf{v}/\hbar c^2}e^{i (\mathbf{v}\cdot{\mathbf{a}})(\mathbf{v}\cdot\mathbf{P})/2\hbar c^2},\label{PBT}
\end{equation}
and accordingly, under the above transformation, the coordinates of the event $(\mathbf{x},t)$ are given by
$(\mathbf{x}',t')=(\mathbf{x}+(\mathbf{v}\cdot\mathbf{a})\mathbf{v}/2c^2,t+\mathbf{v}\cdot\mathbf{a}/c^2)+\mathcal{O}(1/c^4)$. Then the non-relativistic limit of expression (\ref{PBT}) reduces precisely to expression (\ref{EGTBT}) and it produces a relativistic displacement in the coordinates of spacetime. Therefore, the transformation considered in Bargmann's argument produces a non-relativistic phase when acting on the wavefunction of a non-relativistic quantum particle which can be identified as the relativistic time difference between the corresponding inertial observers.

The bottom line is that we should identify $M$ as the generator of time translations of order $1/c^2$ ({\it cf.} r.h.s. of relation (\ref{EGTBT})). And then, the proper description of superposition of states with different masses requires to define an action of the extended Galilei group on the coordinates of Newtoninan spacetime which is consistent with its action on the Hilbert space, in particular, we need to define how $M$ transforms the coordinates of an event such that under the transformation (\ref{EGTBT}), we obtain (at least the time coordinate in) the relativistic relation for $(\mathbf{x}',t')$ up to order $\mathcal{O}(c^0)$. As claimed by Guilini \cite{Giu95}, such description implies that the mass must be dynamical and, accordingly, we have to introduce its conjugate coordinate (the correction of $\mathcal{O}(1/c^2)$ to relativistic time), and then, it may be required to modify the notion of Newtonian spacetime. Such a discussion is presented somewhere else \cite{HC}.

\section*{Acknowledgments}
 
The author is indebted with Y. Bonder, C. Chryssomalakos, E. Okon and D. Sudarsky  for useful discussions. This work has been partially supported by the Czech Ministry of Education, Youth and Sports within the project LC06002.

\end{document}